\documentclass[%
 reprint,
 amsmath,amssymb,
 aps,
 showkeys,
]{revtex4-2}

\usepackage{graphicx}
\usepackage{dcolumn}
\usepackage{bm}
\usepackage{braket}

\begin{document}

\preprint{APS/123-QED}

\title{Experimental storage of photonic polarization entanglement \\ in a broadband loop-based quantum memory}
\author{C.J. Evans}
\author{C.M. Nunn}
\author{S.W.L. Cheng}
\author {J.D. Franson}
\author{T.B. Pittman}
\affiliation{
 Department of Physics, University of Maryland Baltimore County, Baltimore, MD 21250, USA
}%
\date{\today}

\begin{abstract}

We describe an experiment in which one member of a polarization-entangled photon pair is stored in an active ``loop and switch'' type quantum memory device, while the other propagates through a passive optical delay line. A comparison of Bell's inequality tests performed before and after the storage is used to investigate the ability of the memory to maintain entanglement, and demonstrate a rudimentary entanglement distribution protocol. The entangled photons are produced by a conventional Spontaneous Parametric Down Conversion source with center wavelengths at 780~nm and bandwidths of $\sim$10~THz, while the memory has an even wider operational bandwidth that is enabled by the weakly dispersive nature of the Pockels effect used for polarization-insensitive switching in the loop-based quantum memory platform. 
\end{abstract} 

\pacs{XYZ}

\maketitle

A promising approach to long-distance quantum communication involves protocols in which two distant quantum memories become entangled by a central Bell state measurement (BSM) performed on two emitted photons \cite{simon_robust_2003,sangouard_quantum_2011}. The related process in which two entangled photons emitted from a central source are stored in two distant quantum memories enables modified protocols that offer advantages in certain quantum network settings \cite{jones_design_2016}. Recent experimental progress in this direction includes the storage of entangled photons with MHz bandwidths in atomic ensemble quantum memories \cite{choi_mapping_2008,akiba_storage_2009,zhang_preparation_2011,dai_holographic_2012,ding_raman_2015,cao_efficient_2020}, MHz - GHz bandwidths in solid-state quantum memories \cite{clausen_quantum_2011,saglamyurek_broadband_2011,usmani_heralded_2012,bussieres_quantum_2014,zhou_quantum_2015,saglamyurek_quantum_2015,puigibert_entanglement_2020,rakonjac_entanglement_2021,liu_heralded_2021}, and THz bandwidths in a diamond quantum memory \cite{fisher_storage_2017}. In each active approach, the coupling of a photon from a freely propagating mode to the storage mode (typically a collective matter excitation) is accomplished by an externally applied control field that is managed by the user for storage and release of the photons. In addition to bandwidth, key figures of merit include memory efficiency, storage time, accessibility, and output state fidelity. This leads to various trade-offs that can be optimized by different approaches for different applications, and motivates the need for investigations of entanglement storage in other established quantum memory platforms.

Here we investigate entanglement storage in the ``loop and switch'' based quantum memory platform, which offers large bandwidth and high output state fidelity, but relatively low efficiency and discrete time (rather than continuous) accessibility \cite{pittman_cyclical_2002,kaneda_quantum-memory-assisted_2017,bouillard_quantum_2019,takeda_-demand_2019,pang_hybrid_2020,meyer-scott_scalable_2022,hou_entangled-state_2023,arnold_free-space_2023}. In this platform, the coupling between the input/output mode and the storage mode (here, switching into and out of a free-space optical storage loop) is implemented by a user-applied DC control field via the Pockels effect. Because this Pockels effect is only weakly dispersive, loop-based memories of this kind can possess ultra-wide bandwidths that can be matched to that of the entangled photons produced by robust and practical conventional Spontaneous Parametric Down Conversion (SPDC) sources (typically $\sim$ 10~THz). Indeed, broadband Pockels effect based ``loop and switch'' type devices with SPDC sources have recently been used to demonstrate enhanced single-photon production \cite{pittman_single_2002,kaneda_time-multiplexed_2015,kaneda_high-efficiency_2019}, measurement-device-independent quantum key distribution \cite{kaneda_quantum-memory-assisted_2017}, switching and storage of Fock states \cite{bouillard_quantum_2019,alarcon_polarization-independent_2020,leger_amplification_2023}, as well as the generation of multiphoton entangled states \cite{meyer-scott_scalable_2022,hou_entangled-state_2023}. Closely related work includes the manipulation of single-photon streams \cite{istrati_sequential_2020,pang_hybrid_2020,makino_synchronization_2016} and continuous-variable entanglement \cite{takeda_-demand_2019} in loop-based systems.

\begin{figure}[t]
\includegraphics[trim=80 90 170 100, clip, width=3.3in]{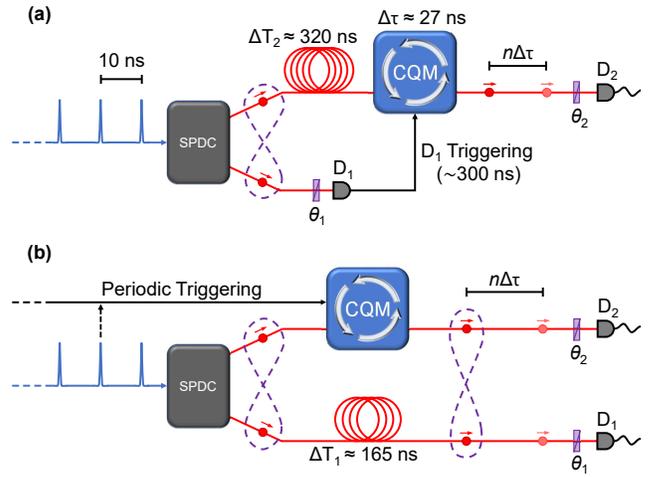}
\caption{Conceptual overview of two types of measurements using the broadband pulsed-SPDC and CQM platform: (a) Initial high data rate alignment and calibration measurements in which the CQM is triggered by the detection of photon 1, and (b) lower data rate entanglement storage measurements in which the CQM is periodically triggered. The red circles denote photons, while the dashed purple lines represents entanglement. $D_{i}$ and $\theta_{i}$ ($i$=1,2) are detectors and polarizers used for various Bell-test measurements.}
\label{fig:overview}
\end{figure}

\begin{figure*}[t]
\includegraphics[trim=145 38 84 20, clip, width=5.5in]{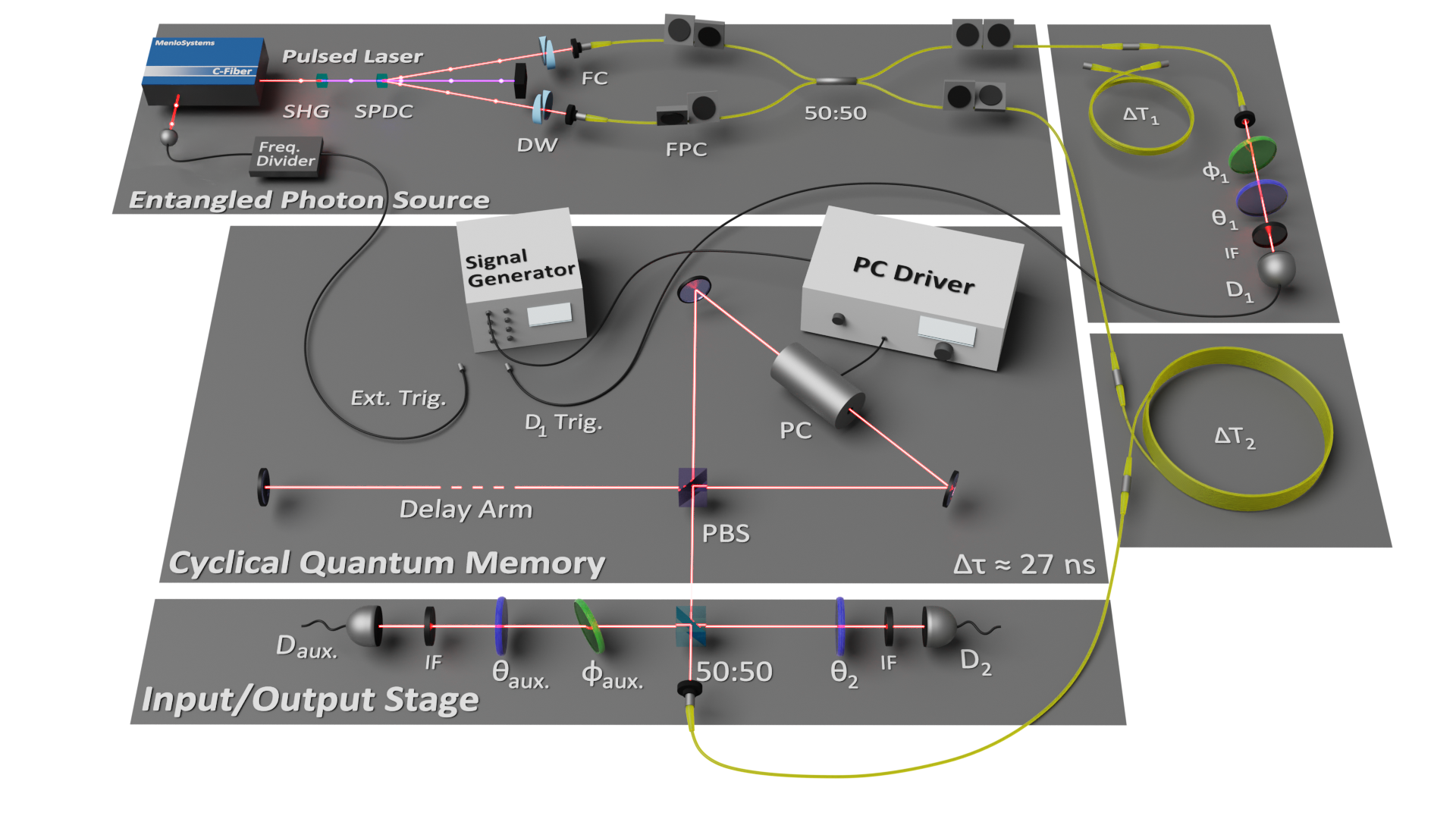}
\caption{Schematic of the experimental apparatus. Translatable delay wedge prisms (DW) are used to adjust the timing of the down-converted photons, and a 50:50 fiber coupler with fiber polarization controllers (FPC) is used to realize entangled states of the form $|\psi^{-}\rangle~=~1/\sqrt{2} (|H_{1}V_{2}\rangle - e^{i \phi} |V_{1}H_{2}\rangle )$ using the Shih-Alley technique \protect\cite{shih_new_1988}. Phase shifters $\phi_{aux}$ and $\phi_{1}$ are used to compensate for the combination of $\phi$ and any net birefringent phase shifts in fiber spools $\Delta T_{1}$, $\Delta T_{2}$, and the CQM itself for the ``before storage'' and ``after storage'' Bell inequality tests. (SHG: second harmonic generation, FC: fiber coupling lenses, IF: 25~nm bandwidth interference filters).}
\label{fig:3D}
\end{figure*}

In the present work on entangled photon storage with this platform, we use the specific loop-based Cyclical Quantum Memory (CQM) geometry of reference \cite{pittman_cyclical_2002}, and a standard SPDC-based source of polarization-entangled photon pairs \cite{shih_new_1988}. The primary experimental challenge is the non-deterministic nature of the SPDC process, which emits the entangled photons at random (and unknown) times. This precludes triggered switching of the photon pairs into the memories, necessitating random attempts at storage and, consequently, low overall data rates in the experiment. These data rates are further reduced by the intrinsic loss in the CQM's ($\sim$22\% per cycle), which currently hinders long-term storage in our setup. In this proof-of-concept demonstration, we overcome these technical problems to some extent by (1) using pulsed SPDC to restrict the possible emission of a photon pair to well-known time intervals, (2) replacing one of the active CQM's with a much lower loss delay line (fiber spool) serving as a ``passive quantum memory'' with a fixed storage time, and (3) performing initial alignment and calibration of the system at higher data rates by triggering the active CQM upon detection of the passively stored photon.

Figure \ref{fig:overview} provides a conceptual overview of these 3 ideas, as well as a summary of the relevant timing parameters and detection system used for Bell-test measurements to verify the stored entanglement. The SPDC source is pumped by a 100~MHz pulse-train, with initial pair production rates on the order of 10~kHz (i.e. an average of 1 pair every 100~$\mu$s), while the CQM has a round-trip cycle time of $\Delta\tau$~$\approx$~27~ns and the experiments involve active storage for up to $n =$ 20~cycles ($\sim$0.5~$\mu$s). Two single photon detectors ($D_{1}$ and $D_{2}$) preceded by polarizers are used to test Bell's inequalities in the system. For the initial testing step shown in Figure \ref{fig:overview}(a), the detection of photon 1 is used to trigger the CQM for storage of photon 2, which is then actively released after $n$ cycles. This preliminary test does not represent the storage of entanglement, but allows us to calibrate the setup at higher data rates. Note in Figure \ref{fig:overview}(a) that photon 2 is delayed by $\Delta T_{2} \sim$ 320~ns to compensate for the latency in the detection and CQM switching process \cite{pittman_cyclical_2002,pittman_single_2002}. 

Next, in Figure \ref{fig:overview}(b), the CQM is periodically triggered by a signal derived from the pump pulse train, which enables the full demonstration of entanglement storage for those cases in which a photon pair is randomly produced at the correct time. In our experiment, we balance the trade-off between a desire for long storage times (i.e. large $n$) and high data rates (i.e. higher frequency triggering) to demonstrate entanglement storage for up to $n =$ 6 cycles ($\sim$162~ns). Note in Figure \ref{fig:overview}(b) that the ``passive quantum memory'' for photon 1 is fixed at a comparable storage time of $\Delta T_{2} \sim$ 165~ns. Note also that difficulties associated with failed storage attempts and loss in the CQM are largely overcome by the post-selective nature of the Bell-tests used to study the ability of the system to store entanglement; only attempts in which both $D_{1}$ and $D_{2}$ register a photon are recorded \cite{clauser_proposed_1969_chsh}.

Figure \ref{fig:3D} shows a schematic of the complete experimental setup. For convenience, the figure highlights five different shaded regions corresponding to the key aspects of the experiment. The SPDC source consists of a 0.7~mm thick BBO crystal pumped by a 100~MHz pulse train at 390~nm derived from the frequency-doubled output of mode-locked fiber laser (Menlo Systems C-Fiber 780; pulse widths $\sim$ 100~fs), and produces photon pairs with central wavelengths of 780~nm. Interference filters with a bandwidth of 25~nm are used to define the photon bandwidths ( $\sim$10~THz). We use Type-I non-collinear SPDC and the Shih-Alley (SA) technique at a 50/50 beamsplitter \cite{shih_new_1988} to post-select entangled states of the form $|\psi^{-}\rangle = 1/\sqrt{2} (|H_{1}V_{2}\rangle - |V_{1}H_{2}\rangle )$, where $H$ and $V$ denote horizontally and vertically polarized photons and the subscripts correspond to output channels 1 and 2. Photon 1 is detected in the direct output of the SA beamsplitter, while photon 2 is sent to the CQM which, in our laboratory, is located on a second optical table roughly 6~m from the source.

Complete details of the operational technique of the loop-based CQM are provided in reference \cite{pittman_cyclical_2002}. To summarize, it consists of a high-speed Pockels cell (PC) placed in a Sagnac-like interferometer formed by a polarizing beamsplitter (PBS), two broadband mirrors, and a lengthy ``out and back'' delay arm. Incident photons are delocalized into two counter propagating $H$ and $V$ polarization components that are repeatedly ``flipped'' (i.e. $H \leftrightarrow V$) each time they pass through the PC in the ``on'' state. This leads to a self-cancellation effect for phase shift errors due to birefringence in the CQM for photons stored for an even number $n$ of cycles. In addition, bit flip errors (imperfect polarization rotations) are ejected from the CQM geometry at incorrect times, contributing only to overall loss. 

We use a Lithium Tantalate (LTA) multi-crystal based PC operated in a transverse configuration (ConOptics model 360-80, with 25D driver), with a half-wave voltage of only 140~V at 780~nm, and rise-time and fall times (i.e. switching times) of $\sim$15~ns, which are safely shorter than the 27~ns cycle time of the CQM. The user actively stores the incident photon, and releases it after a chosen value of $n$ cycles, by simply switching the PC between its ``on'' and ``off'' states at the appropriate times. 

As shown in Figure \ref{fig:3D}, the PC driver is activated by a short-pulse signal generator that is triggered by either (1) the detection of a photon in $D_{1}$ for the initial tests of Figure \ref{fig:overview}(a), or (2) a periodic signal derived by frequency division of a 100~MHz synchronization signal from the mode-locked laser for the main entanglement storage experiments of Figure \ref{fig:overview}(b). In addition, the delays $\Delta T_{1}$ and $\Delta T_{2}$ required for these two types of experiments are formed by fiber spools that can be inserted and removed as needed.

A key 50/50 beamsplitter is inserted in the CQM input channel to reflect the CQM output to the second Bell test detector, $D_{2}$. While this reduces the CQM overall efficiency to a maximum value of 25$\%$ in this proof-of-concept experiment, it also provides a valuable auxiliary detection channel that can be used in-situ for comparative Bell tests: Bell tests ``before storage'' use correlations between detectors $D_{1}$ and $D_{aux}$, while Bell tests ``after storage'' are performed with $D_{1}$ and $D_{2}$. We note that this efficiency limitation can be overcome by replacing the 50:50 beamsplitter with a high-quality optical circulator or polarization to time-bin transduction methods in the loop-based memory platform \cite{kaneda_quantum-memory-assisted_2017,arnold_free-space_2023,meyer-scott_scalable_2022}.

Additional tests of the PC with various lasers confirmed that with the half-wave voltage set for perfect 90$^{o}$ polarization rotation (``flipping'') of 780~nm light, a wavelength range greater than 25~nm (near 780~nm) would would be ``flipped" with greater than 95$\%$ fidelity due to the weakly dispersive birefringence of the Pockels effect in our system. Combined with very broadband CQM mirror reflectivities ($R > 98~\%$ for $H$ and $V$ components over 750~nm - 1100~nm), this enables the CQM to serve as a high-speed broadband optical quantum memory device.

\begin{figure}[t]
\includegraphics[trim=189 110 190 90, clip, width=3.35in]{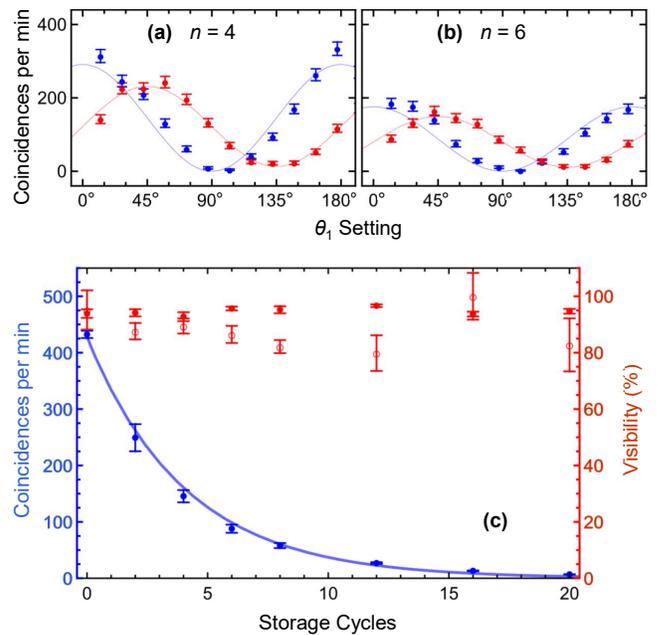}
\caption{Summary of experimental results from the high data rate test and alignment measurements: (a) heralded output state polarization measurements for storage of $n$~=~4 cycles. Blue data corresponds to $\theta_{2}$ fixed at $0^{o}$, while red data corresponds to $\theta_{2}~=~45^{o}$; (b) analogous results for $n$~=~6 cycles; (c) summary of extracted average coincidence rates (blue data) and visibilities in the $-45^{o}/45^{o}$ basis from ``before storage'' runs (red data, open circles) and ``after storage'' runs (red data, closed circles) for various storage times up to $n$~=~20 cycles (540~ns). }
\label{fig:prelim}
\end{figure}

Figure \ref{fig:prelim} shows a summary of calibration data using the arrangement of Figure \ref{fig:overview}(a). While this arrangement does not demonstrate the storage of entanglement, we perform various measurements using the same polarizer settings needed to characterize the expected performance of the system for subsequent Bell tests. Figures \ref{fig:prelim}(a) and \ref{fig:prelim}(b) show plots of the coincidence counting rates between $D_{1}$ and $D_{2}$ as a function of $\theta_{1}$ after heralded storage of photon 2 for $n$~=~4 and $n$~=~6 cycles, respectively, for the cases of $\theta_{2}$ fixed at $0^{o}$ (blue data) and $45^{o}$ (red data). The sinusoidal fits to the data are then used to extract the visibilities from the red curves (a measure of output state polarization fidelity) and the average counting rates from the blue curves (a measure of loss during storage). Analogous data sets (not shown) were taken for storage up to $n$~=~20 cycles, as well as for the ``before storage'' case using coincidence counts between detectors $D_{1}$ and $D_{aux}$. These data sets were all taken in rapid succession with minimal adjustments to the setup in order to study the system's performance as a function of $n$. Figure \ref{fig:prelim}(c) summarizes this data and provides two main results: an exponential fit to the average count rate (blue data) shows a CQM loss of roughly 22$\%$ per cycle \cite{LossFootnote}, while the ``after storage'' visibility (red data, closed circles) shows essentially no degradation with increasing storage time. 

\begin{figure}[t]
\includegraphics[trim=196 171 190 176, clip, width=3.4in]{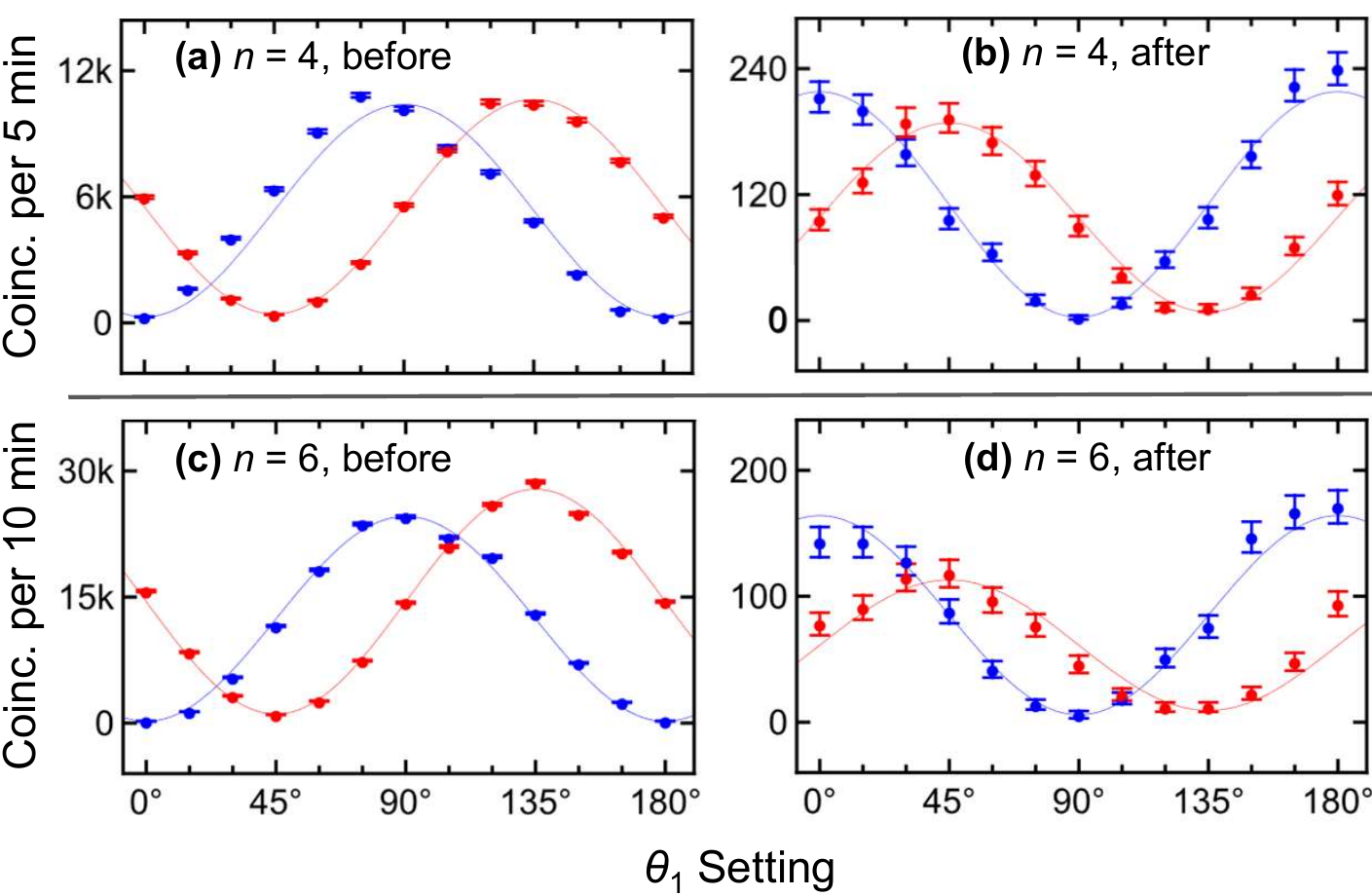}
\caption{Summary of experimental results demonstrating entanglement storage: (a) and (b) show ``before storage'' and ``after storage'' polarization correlations for the case of $n$~=~4 cycles, with blue (red) data corresponding to $\theta_2$ fixed at $0^{\circ}$ ($45^{\circ}$). The sinusoidal data in (b) is shifted by $90^{\circ}$ due to the intrinsic bit flipping in the CQM. (c) and (d) show analogous results for the case of $n$~=~6 cycles. The high visibilities of the fits to the data violate the CHSH form of Bell's inequality \cite{tittel_long-distance_1999} and demonstrate the ability to store entanglement in the SPDC and loop-based quantum memory platform.}
\label{fig:mainData}
\end{figure}

These preliminary loss and visibility results of Figure \ref{fig:prelim}(c) provide an expectation of being able to violate Bell's inequality after storing entanglement using the arrangement of Figure \ref{fig:overview}(b) for increasingly long storage times, until the overall loss in the CQM drives the signal-to-noise ratio in the system down to an unmanageable level. This leads to the main results of the paper illustrated in Figure \ref{fig:mainData}, which show examples of this entanglement storage for the case of $n$~=~4 and 6 cycles.

Figures \ref{fig:mainData}(a) and \ref{fig:mainData}(b) correspond to the ``before storage'' and ``after storage'' coincidence count data for the case of $n$~=~4 cycles. The experimental data shows the expected sin$^{2}(\theta_{1} - \theta_{2})$ signature of the $|\psi^{-}\rangle$ Bell state, with``before storage'' measured visibilities of $(95~\pm~3)\%$ in the $H/V$ basis and $(92~\pm~1)\%$ in the $-45^{o}/45^{o}$ basis, and corresponding ``after storage'' visibilities of $(97~\pm~1)\%$ and $(91~\pm~4)\%$. As is well known, combined visibilities greater than 71\% in these experimental situations are sufficient for a violation of the CHSH form of Bell’s inequality subject to certain reasonable assumptions \cite{clauser_proposed_1969_chsh}, and here correspond to Bell parameter values of $S~=~2.64~\pm~0.04$ before storage, and $S~=~2.66~\pm~0.06$ after storage \cite{tittel_long-distance_1999}. In this proof-of-concept experiment, these $S~>~2$ parameter values provide a demonstration of the ability to store and maintain polarization entanglement in the loop-based quantum memory platform.

Figures \ref{fig:mainData}(c) and \ref{fig:mainData}(d) show analogous results for the case of $n$~=~6 cycles. Here the ``before storage'' measured visibilities are $(98~\pm~1)\%$ in the $H/V$ basis and $(93~\pm~1)\%$ in the $-45^{o}/45^{o}$ basis, with corresponding ``after storage'' visibilities of $(93~\pm~3)\%$ and $(85~\pm~7)\%$. These visibilities correspond to Bell parameter values of $S = 2.69 \pm 0.02$ before storage and $S = 2.52 \pm 0.11$ after storage, once again demonstrating successful entanglement storage. We suspect the slightly lower $S$ value in the ``after storage" case was primarily due to small misalignments that occurred in the CQM between data runs, as well as the reduction in overall data rates which necessitated longer collection times and thus increased experimental instabilities during the $n$~=~6 cycle run. As a technical point of interest in the experiment, the longer storage times associated with increasing values of $n$ require larger divisions of the 100 MHz periodic CQM triggering signal to prevent accidental “on/off” state overlap in the PC during storage. These less frequent attempts at storage corresponded to significant data rate reductions for large $n$, which are then further reduced by the accumulating intrinsic CQM loss of 22$\%$ per cycle. An accurately calibrated observation of these count rate reductions is not possible in Figure \ref{fig:mainData} due to small changes in the experimental conditions that occurred between these non-sequential data runs. Nonetheless, further increases in these two types of data rate reduction mechanisms for increasing $n$ prevented realistic attempts at entanglement storage for, say, 20 cycles in our current proof-of-concept type setup.

In summary, we have demonstrated the storage of entangled photons with $\sim$10~THz bandwidths from a conventional SPDC source using one active broadband loop-based quantum memory device, and a second ``passive quantum memory'' formed by a simple delay line, in analogy with earlier entanglement storage demonstrations using other quantum memory platforms \cite{akiba_storage_2009,clausen_quantum_2011,saglamyurek_broadband_2011,saglamyurek_quantum_2015,bussieres_quantum_2014,zhou_quantum_2015,fisher_storage_2017}. The experimental results represent a demonstration of a rudimentary entanglement distribution protocol, in which one member of an entangled pair is delivered to a location $A$ at a fixed time, while the other is delivered to a second distant location $B$ at an arbitrarily chosen time. 

The broadband nature of the loop-based memory platform helps overcome the notorious ``bandwidth matching'' problem associated with using traditional broadband SPDC entangled photon sources and narrowband atomic quantum memories \cite{akiba_storage_2009}. However, it is important to note that the lack on an intrinsic optical nonlinearity the loop-based platform represents a drawback for multi-node quantum repeater type applications in which a single node acts as both a memory and a quantum processor \cite{krutyanskiy_telecom-wavelength_2023}. For these more challenging protocols, supplementing the loop-based memory with additional probabilistic techniques from the linear optics quantum computing paradigm would be required \cite{knill_scheme_2001}. 

The primary limitation in this proof-of-concept experiment was the use of randomly produced entangled photon pairs, which essentially necessitated random attempts at storage and thus low overall data rates. For future applications, these difficulties can be completely overcome by the use of heralded entangled pairs that can be produced by combining several random SPDC sources \cite{sliwa_conditional_2003,pittman_heralded_2003,wagenknecht_experimental_2010,barz_heralded_2010}. We note that for more demanding applications, these same heralded SPDC sources can, in principle, be converted to ``on-demand'' entanglement sources by using two loop-based memories and some of the techniques demonstrated in the present work \cite{pittman_heralded_2003}. Consequently, near-term implementations of various multi-photon quantum networking and entanglement distribution protocols could benefit from the use of robust broadband SPDC sources and ultra-broadband loop-based memories, and the proof-of-concept experimental results presented here represent a tangible step in that direction.

{\bf Acknowledgements:} This work was supported by the National Science Foundation under Grant No. 2013464.

\newpage

\

\bibliography{mainRev}

\end{document}